\documentclass[12pt,a4paper]{article}

\DeclareFontFamily{OT1}{rsfs}{}
\DeclareFontShape{OT1}{rsfs}{m}{n}{ <-7> rsfs5 <7-10> rsfs7 <10->rsfs10}{} \DeclareMathAlphabet{\mycal}{OT1}{rsfs}{m}{n}

\newcommand{\be}{\begin{equation}}
\newcommand{\ee}{\end{equation}}
\newcommand{\beq}{\begin{eqnarray}}
\newcommand{\eeq}{\end{eqnarray}}

\usepackage{amsmath,amsthm,amsfonts}
\input amssym.def
\newsymbol\square 1003
\begin{document}

\centerline{\bf  $SO(n+1)$  SYMMETRIC SOLUTIONS OF THE EINSTEIN}
\centerline{\bf  EQUATIONS IN HIGHER DIMENSIONS}
\bigskip
\centerline{M. Jakimowicz and J. Tafel}

\null

\centerline{Institute of Theoretical Physics, University of Warsaw,}
\centerline{Ho\.za 69, 00-681 Warsaw, Poland, email: tafel@fuw.edu.pl}

\bigskip
\bigskip
\begin{abstract}
A method of solving the Einstein equations with a scalar field is presented. It is applied to  find higher dimensional vacuum metrics  invariant under the group $SO(n+1)$ acting on n-dimensional spheres.
\end{abstract}

PACS number: 04.50.-h

\section{Introduction}
Recently, higher dimensional solutions of the Einstein equations became important because of  a great interest in  string theories and induced effective theories in 4+d dimensions (see e.g. \cite{M} and references therein). In the  brane-world  gravity  matter fields are  usually confined to a 4-dimensional brane and gravity can propagate in extra dimensions. At the classical level the gravitational field of a bulk should satisfy the vacuum Einstein equations, possibly with a cosmological constant. In order to find and classify higher dimensional solutions methods of the standard general relativity were generalized (see \cite{C,PPO,OPZ} and references therein). 

One of the most effective techniques of solving the Einstein equations is their reduction via symmetries. Taking into account a role of the Birkhoff theorem in general relativity it is not surprising that much attention is payed to higher dimensional metrics admitting symmetries of 2-dimensional or higher dimensional sphere. For instance, this property is shared by the 5-dimensional Gross-Perry metric \cite{GP} studied in the  framework of the Kaluza-Klein theory (see \cite{PL} for a recent discussion of  this metric). 
Recently, a more systematic investigation of 5-dimensional $SO(3)$ symmetric vacuum metrics  was  performed by Lake \cite{L} and Millward \cite{Mil}. 
 
In this paper we consider $(n+N+1)$- dimensional metrics invariant under the rotation group $SO(n+1)$ acting on n-dimensional spheres. Following the Kaluza-Klein approach (see e.g. \cite{CJ}) we first recall the dimensional reduction of the Einstein equations to equations in $N+1$ dimensions with a scalar field $\phi$ and an exponential potential. In section 3 we reduce the latter equations, with a general potential, under the assumption that surfaces $\phi=const$ are Einstein spaces and their normal vector field is geodetic. We obtain a closed system of two ordinary differential equations (they correspond to the Friedmann equations in cosmology)  and  algebro-geometric conditions on the metric of the surfaces.  For $N=n=2$ they describe a class of metrics which generalizes those considered in \cite {L,Mil}.
  Finally, we present examples of vacuum metrics obtained by our method. Among them there are new solutions which belong to the generalized Kundt class \cite{CMPP}. 
  
\section{Symmetry reduction of the Einstein equations with a cosmological constant}
Let M be a $(n+N+1)$--dimensional manifold with a Lorentzian metric $g$ admitting $SO(n+1)$
spherical symmetry. We assume that orbits of the group are diffeomorphic to the n-dimensional sphere $S_n$. In local coordinates
$x^{\mu}=\left\{x^a,x^A\right\}$,
 $a=0,1,\cdots,N;$ $A=N+1,\cdots,N+n$, the metric can be written in the following form
\begin{equation} \label{1}
g=g_{ab}dx^adx^b-e^{2f}s_{AB}dx^Adx^B\ ,
\end{equation}
where $s_{AB}dx^Adx^B$ is the standard metric of $S_n$ 
and $g_{ab}$ and $f$ are functions of coordinates  $x^a$ (note that for n=1 (\ref{1}) is not the most general invariant metric). 

Components of the Ricci tensor of (\ref{1}) read 
\begin{eqnarray}
\label{5}
R_{aB}&=&0\\
\label{8}
R_{AB}&=&\left(n-1+\square' f+nf^{|a}f_{|a}\right)s_{AB}\\
\label{9}
R_{ab}&=&R{'}_{ab}-nf_{|a}f_{|b}-nf_{|ab}\ .
\end{eqnarray}
Here $_{|a}$, $R^{'}_{ab}$  and $\square'$ denote, respectively, the covariant derivative, the
Ricci tensor and  the d'Alembert operator of metric
$g'=g_{ab}dx^adx^b$.

The vacuum Einstein equations with a cosmological constant $\Lambda$ in $(n+N+1)$ dimensions are equivalent to
\begin{equation}\label{8a}
R_{\mu\nu}=\frac{2\Lambda}{1-N-n} g_{\mu\nu}\ .
\end{equation}
For $N=1$ solutions of (\ref{8a}) of the form (\ref{1}) are multidimensional Schwarzschild-de Sitter metrics 
\begin{equation}\label{8b}
g=\left(1-\frac{2M}{r^{n-1}}-\frac{2\Lambda r^2}{n\left(n+1\right)}\right)dt^2-\left(1
-\frac{2M}{r^{n-1}}-\frac{2\Lambda r^2}{n\left(n+1\right)}\right)^{-1}dr^2-r^2s_{AB}dx^Adx^B.
\end{equation}
In what follows we assume $N>1$. In this case we can apply to $g^{'}$ a conformal transformation  of the form
\begin{equation}\label{10}
\widetilde{g}_{ab}=e^{\frac{2n}{N-1}f}g_{ab}
\end{equation}
which induces the following changes
\begin{equation}\label{11}
\widetilde{R}_{ab}=R'_{ab}-nf_{|ab}+\frac{n^2}{N-1}f_{|a}f_{|b}-
\frac{n}{N-1}(\square'f+nf^{|c}f_{|c})g_{ab}
\end{equation}
\begin{equation}\label{12}
\widetilde{\square}f=e^{-\frac{2n}{N-1}f}(\square'f+nf^{|c}f_{|c})\ .
\end{equation}
Here $\widetilde{\square}f=\widetilde{g}^{ab}f_{;ab}$ and the
semicolon denotes the covariant derivative with respect to metric
$\widetilde{g}$.

By virtue of (\ref{11}) and (\ref{12}) expresions (\ref{8}), (\ref{9}) take the form
\begin{eqnarray}
\label{13}
R_{AB}&=&\left(n-1+e^{2(\frac{n+N-1}{N-1}f)}\widetilde{\square}f\right)s_{AB}\\
\label{14}
R_{ab}&=&\widetilde{R}_{ab}-n\frac{n+N-1}{N-1}f_{;a}f_{;b}+\frac{n}{N-1}\widetilde{\square}f
\widetilde{g}_{ab}
\end{eqnarray}
and equations (\ref{8a}) reduce to 
\begin{equation} \label{15}
\widetilde{\square}f=-\left(n-1\right)e^{-2\frac{n+N-1}{N-1}f}+\frac{2\Lambda}{n+N-1} e^{-\frac{2n}{N-1}f}
\end{equation}
and 
\begin{equation}\label{16}
\widetilde{R}_{ab}=n\frac{n+N-1}{N-1}f_{;a}f_{;b}-\left(\frac{n}{N-1}\widetilde{\square}f
+\frac{2\Lambda }{n+N-1}e^{-\frac{2n}{N-1}f}\right)\widetilde{g}_{ab}\ .
\end{equation}
We substitute  (\ref{15}) into (\ref{16}) 
 and we pass to the Einstein tensor $\widetilde{G}_{ab}$ of $\widetilde{g}$. In this way we obtain
\begin{equation} \label{17}
\widetilde{G}_{ab}=n\frac{n+N-1}{N-1}\left(f_{;a}f_{;b}-\frac{1}{2}f^{;c}f_{;c}\widetilde{g}_{ab}\right)+\left(-\frac{1}{2}n\left(n-1\right)e^{-2\frac{n+N-1}{N-1}f}+\Lambda
e^{-\frac{2n}{N-1}f}\right)\widetilde{g}_{ab}.
\end{equation}
After rescaling
\begin{equation}\label{18}
\phi=\sqrt{\frac{n\left(n+N-1\right)}{\left(N-1\right)}}f
\end{equation}
equations (\ref{15}) and (\ref{17}) take the form of $(N+1)$--dimensional
Einstein equations with the scalar field $\phi$
\begin{equation}\label{19}
\widetilde{G}_{ab}=\phi_{;a}\phi_{;b}+ \left(V(\phi)-\frac{1}{2}\phi^{;c}\phi_{;c}\right)\widetilde{g}_{ab}
\end{equation}
\begin{equation}\label{20}
\widetilde{\square}\phi=-V_{,\phi}\ .
\end{equation}
The potential $V$ is given by 
\begin{equation}\label{21}
V=-\frac{1}{2}n\left(n-1\right)e^{-2\sqrt{\frac{n+N-1}{n\left(N-1\right)}}\phi}+\Lambda
e^{-2
\sqrt{\frac{n}{\left(N-1\right)\left(n+N-1\right)}}\phi}\ .
\end{equation}
 For a non constant function 
$\phi$ equation (\ref{20}) follows from (\ref{19}) and it is equivalent to the 
energy-momentum conservation law $T^{ab}_{\ \ ;b}=0$.

\section{Reduction of the Einstein equations with a scalar field}

\bigskip

Let us consider equations (\ref{19}), (\ref{20}) in  spacetime of dimension $N+1\geq 3$.
Assume that $\phi_{,a}\neq 0$ and  surfaces $\phi$=const are not null. Then
there are coordinates
$\phi,x^i$ such that 
\be
\tilde g=\tilde g_{\phi\phi}d\phi^2+\tilde g_{ij}dx^idx^j\ .\label{3.3}
\ee
The coordinate $\phi$ is timelike if $\tilde g_{\phi\phi}>0$. In this case we set $x^0=\phi$
and $i=1,...N$. If $\tilde g_{\phi\phi}<0$ $\phi$ is spacelike, $x^N=\phi$ and $i=0,..,N-1$.

Assume moreover that $\tilde g_{\phi\phi}$ is independent of
$x^i$. Then we can find a new coordinate
$s$  such that 
\be
\phi=\phi(s)\label{3.4a}
\ee
 and 
\be
\tilde g=\epsilon ds^2+\tilde g_{ij}dx^idx^j\ ,\ \epsilon=\pm 1\ .\label{3.4}
\ee
Geometrically, above assumptions mean that the field of normal vectors to
surfaces $\phi$=const is geodesic, timelike or spacelike, and $s$ is the
affine parameter along the field.

Under conditions (\ref{3.4a}), (\ref{3.4}) equation (\ref{20}) yields
\be
\ddot{\phi}+\dot \phi(\ln{\sigma})\dot{\phantom{l}}=-\epsilon V_{,\phi}\ ,\label{3.12}
\ee
where
\be
\sigma=|det(\tilde g_{ij})|^{\frac{1}{2}}\label{3.8}
\ee
and the dot denotes the partial derivative with respect to $s$.
It follows from (\ref{3.12}) that 
\be 
\sigma=\beta(s)\sigma_0(x^i)\label{3.13}
\ee
and
\be
(\beta \dot \phi)\dot{\phantom{l}}=-\epsilon \beta V_{,\phi}\ ,\label{3.14}
\ee
where $\beta$ is a function independent of coordinates $x^i$ and $\sigma_0$ is
independent of $s$.

The Einstein tensor of metric (\ref{3.4}) takes the following form
\be
\tilde G^{\phi}_{\ \phi}=-\frac{1}{2}\hat R +\epsilon \sigma^{-2}\Pi\label{3.5}
\ee
\be
\tilde G^{\phi}_{\ j}=\epsilon (\sigma^{-1}\pi^k_{\ j})_{;k}\label{3.6}
\ee
\be
\tilde G^i_{\ j}=\hat G^i_{\ j}-\epsilon\sigma^{-1}\dot \pi^i_{\ j}-\epsilon \sigma^{-2}\Pi\delta^i_{\
j}\label{3.7}
\ee
where
\be
\Pi=\frac{1}{2}\big[\frac{(\pi^i_{\ i})^2}{N-1}-\pi^i_{\ j}\pi^j_{\ i}\big]\ ,\label{3.9}
\ee
quantities 
\be
\pi^i_{\ j}=\frac{1}{2}\sigma \tilde g^{ik}\dot{\tilde g}_{kj}-\dot \sigma \delta^i_{\ j}\label{3.10}
\ee
are related to the exterior curvature of surfaces $s$=const
 and $\hat G^i_{\ j}$
and $\hat R$ are, respectively, the Einstein tensor and the Ricci scalar of the 
metric 
\be
\hat g=\tilde g_{ij}dx^i dx^j.\label{3.10a}
\ee

In order to simplify the r.h.s. of (\ref{3.7}) let us assume that 
\be 
\hat G^i_{\ j}=\hat \Lambda \delta^i_{\ j}\label{3.11}
\ee
(note that $\hat \Lambda=0$ if $N=2$). It
follows from (\ref{3.11}) and the Bianchi identities that $\hat \Lambda=\hat
\Lambda (s)$.
Equations (\ref{19}) with indices $i,j$ now yield
\be
\dot\pi^i_{\ j}= \sigma(\epsilon\hat \Lambda-\sigma^{-2}\Pi+\frac{1}{2}\dot
\phi^2-\epsilon V)\delta^i_{\ j}\ .\label{3.15}
\ee
Since $\dot\pi^i_{\ j}\sim \delta^i_{\ j}$ the matrix $\pi^i_{\ j}$ must have the following structure
\be 
\pi^i_{\ j}=a \delta^i_{\ j}+P'^i_{\ j}(x^k)\ ,\ \ P'^i_{\ \ i}=0.\label{3.16}
\ee
The function $a$ can be related to $\beta$ and $\sigma_0$ by substituting (\ref{3.16}) and
(\ref{3.13}) into the identity 
\be
\pi^i_{\ i}=(1-N)\dot \sigma\label{3.16a}
\ee
 which follows
from (\ref{3.10}). Consecutively we obtain
\be 
\pi^i_{\ j}=\sigma_0[(\frac{1}{N}-1)\dot \beta\delta^i_{\ j}+P^i_{\ j}(x^k)]\ ,\label{3.17}
\ee
where
\be
\ \ P^i_{\ i}=0\label{3.17a}
\ee
and
 the matrix $P=(P^i_{\ j})$ is independent of $s$.
Substituting (\ref{3.17}) back to (\ref{3.15}) yields a condition on 
$P$
\be
P^i_{\ j}P^j_{\ i}=2c=const\label{3.18}
\ee
and the following equation for the functions $\beta(s)$
and $\phi(s)$
\be 
(\frac{1}{N}-1)\big(\frac{\ddot{\beta}}{\beta}-\frac{\dot\beta^2}{2\beta^2}\big)-\frac{c}{\beta^2}
-\frac{1}{2}\dot\phi^2+\epsilon (V-\hat\Lambda)=0\ .\label{3.19}
\ee
Given (\ref{3.17}) and (\ref{3.13})
relation (\ref{3.10}) becomes a linear equation for the
matrix $\hat g=(\tilde g_{ij})$ (if there is no confusion we denote a metric and the
corresponding matrix of its components by the same symbol). Its general solution has the form
\be
\hat g=\beta^{2/N}\gamma (x^i)e^{P\tau(s)}\ ,\label{3.20}
\ee
where $\gamma=(\gamma_{ij}) $ is a nondegenerate matrix
independent of $s$ and function $\tau$ is related to $\beta$ via
\be
\beta\dot\tau=2.\label{3.21}
\ee
In order to guarantee that the r.h.s. of (\ref{3.20}) is a symmetric matrix we have to
require 
\be
\gamma_{ij}=\gamma_{ji}\ ,P_{ij}=P_{ji}\ ,\label{3.22}
\ee
where $P_{ij}=\gamma_{ik}P^k_{\ j}$. Note that equation (\ref{3.20}) implies (\ref{3.13}) with $\sigma_0=|det
\gamma|^{1/2}$.

Thus, under assumptions (\ref{3.4a}), (\ref{3.4}) and
(\ref{3.11}) equation (\ref{20}) takes the form (\ref{3.14}) and equations
(\ref{19}) with indices $i,j$ are equivalent to (\ref{3.17a})-(\ref{3.22}).

Let us consider now equations (\ref{19}) with indices $a,\phi$. From the point of view
of an evolution with respect to the coordinate $s$ these equations are
constraints. For indices $i,\phi$ they take the form
\be
P^k_{\ i;k}=0\ .\label{3.25}
\ee
Due to (\ref{3.17a}) and (\ref{3.18}) the $s$ derivative of the l.h.s. of
(\ref{3.25}) vanishes.  Hence, it is sufficient to solve (\ref{3.25})
with covariant derivatives defined by the $s$-independent metric $\gamma$.

By virtue of (\ref{3.5}), (\ref{3.9}), (\ref{3.17})-(\ref{3.18}) equation (\ref{19}) with indices $\phi,\phi$ is equivalent to 
\be 
\epsilon\hat R=(1-\frac{1}{N})\frac{\dot\beta^2}{\beta^2}-\frac{2c}{\beta^2}
-\dot\phi^2-2\epsilon V\ .\label{3.23}
\ee
It follows from (\ref{3.23}) that $\hat R$ cannot depend on coordinates $x^i$.
Taking the $s$ derivative of (\ref{3.23}) and comparing with (\ref{3.19})
yields
\be 
\beta^{2/N}\hat R=const\ .\label{3.24}
\ee
Equations (\ref{3.11}) and (\ref{3.24}) can be jointly
written as the following condition on the Ricci tensor of the metric $\gamma e^{P\tau}$ 
\be 
R^i_{\ j}(\gamma e^{P\tau})=\lambda\delta^i_{\ j}\ ,\ \ \lambda=const\ .\label{3.24a}
\ee

Due to (\ref{3.24a}) equation (\ref{3.23}) is the first integral of
(\ref{3.19}) and  equation (\ref{3.19}) 
can be postponed if $\dot \beta\neq 0$.
If $\beta$=const equations  (\ref{3.14}), (\ref{3.19}), (\ref{3.21}) and
(\ref{3.23}) admit solutions only if $V$=const. In this case, without a loss of generality, we
can assume that
\be
\beta=1\ ,\ \ V=-\frac{1}{2}(N-1)\lambda\label{3.24b}
\ee
\be
\tau=2s\ ,\ \ \phi=s\sqrt{-2c-\epsilon\lambda}\ ,\ \ 2c<-\epsilon\lambda\ .\label{3.24c}
\ee

Summarizing, in order to construct a class of solutions of the Einstein equations with
a scalar field and a nonconstant potential $V$ we can proceed along the following steps:
\begin{itemize}
\item
Find N-dimensional metric $\gamma_{ij}$ and a symmetric tensor $P_{ij}$ such
that conditions (\ref{3.24a}), (\ref{3.17a}), (\ref{3.18}) and (\ref{3.25}) (with covariant derivatives defined by $\gamma$) are satisfied.
\item
Find solutions $\phi$, $\beta\neq$ const of ordinary differential equations 
(\ref{3.14}), (\ref{3.23}).
\item
Construct (N+1)-dimensional metric according to (\ref{3.4}), (\ref{3.20}) and
(\ref{3.21}).
\end{itemize}
It follows from the reduction in section 2 that for $V$  given by  (\ref{21}) metric $\tilde g$ and the scalar field $\phi$ define a ($n+N+1)$-dimensional Einstein metric. In this case $g_{ab}$ and $f$ are given by (\ref{10}) and (\ref{18}).

\section{Examples}
For any dimension $N>1$ the conditions on $\gamma$ and $P$ from section 3 are obviously fulfilled by 
\be
\gamma_{ij}=\gamma_{ji}=const\ ,\ \ P_{ij}=P_{ji}=const\ ,\ \ P^i_{\ i}=0\ .\label{3.25a}
\ee
For $\epsilon=1$ (\ref{3.25a}) yields the Misner parametrization \cite{Mis} of the Bianchi I cosmological models. In this case we can assume that $\gamma_{ij}=-\delta_{ij}$ and $P$ is diagonal. For $\epsilon=-1$ we can transform $\gamma$ into the N-dimensional Minkowski metric.
In this case the matrix $P$ can be simplified by means of N-dimensional Lorentz transformations. For N=2 one obtains  the following canonical forms of the metric $\gamma e^{P\tau}$
\be
e^{a\tau}dt^2-e^{-a\tau}dx^2\label{d1}
\ee
\be
\cos{(a\tau)}(dt^2-dx^2)+2\sin{(a\tau)}dtdx\label{d2}
\ee
\be
du(dv+a\tau du).\label{d3}
\ee
For N=3 they read 
\be
e^{a\tau}dt^2-e^{b\tau}dx^2-e^{-(a+b)\tau}dx^2\label{d4}
\ee
\be
e^{b\tau}\cos{(a\tau)}(dt^2-dx^2)+2e^{b\tau}\sin{(a\tau)}dtdx-e^{-2b\tau}dy^2\label{d5}
\ee
\be
du(e^{b\tau}dv+a\tau du)-e^{-2b\tau}dy^2.\label{d6}
\ee

Here $t,\ x$, or $u$, $v$, and $y$ are coordinates and $a,\ b$ are constants. To simplify $P$ and $\gamma e^{P\tau}$ for $N>3$ a classification of symmetric tensors in Lorentz manifolds can be usefull (see \cite{Kr,MCPP} and references therein). Metrics considered in \cite{GP,L,Mil} are related to particular  realizations of (\ref{d1}).

In the case $N=2$  we can find general solution conditions for  $\gamma$ and $P$. Indeed, for $N=2$ equation (\ref{3.11}) is identically satisfied with
$\hat\Lambda=0$.
For any $N$ the $s$ derivative of the l.h.s. of
(\ref{3.24}) vanishes by virtue of (\ref{3.17a}) and (\ref{3.25}). Hence, for $N=2$ condition
(\ref{3.24a}) is equivalent 
to the requirement that metric $\gamma_{ij}$ has a constant curvature. For instance, if $\epsilon=-1$,  $\gamma$ reads
\be 
\gamma=\frac{dudv}{(1+\frac{\lambda}{4} uv)^2}\ ,\label{3.34}
\ee
where $u,\ v$ are  null coordinates and $\lambda$ is a constant. 
Given (\ref{3.34}) conditions 
 (\ref{3.17a}), (\ref{3.18}) and (\ref{3.25})  can be fully solved. If $c\neq 0$ then $\lambda=0$ and $\gamma e^{P\tau}$ is given by (\ref{d1}) or (\ref{d2}).

 If  $c=0$ one obtains
\be
\gamma e^{P\tau}=\frac{dudv}{(1+\frac{\lambda}{4} uv)^2}+\tau h(u)du^2\ ,\label{3.35}
\ee
where $h$ is an arbitrary function of $u$. Note that (\ref{3.35}) leads to  vacuum metrics (\ref{1}), which belong to the generalized Kundt class \cite{CMPP}.

 Let us consider equations (\ref{3.14}) and (\ref{3.23}) with $\epsilon=-1$ and the potential $V$ given by (\ref{21}) with $\Lambda=0$. 
For $n=1$ $V=0$ and these equations can be  solved up to quadratures since 
equations (\ref{3.14}) and  (\ref{3.21}) imply
\be
\dot{\phi}=\frac{2c'}{\beta}\ ,\ c'=const\label{3.35a}
\ee
\be
\phi=c'\tau\label{3.35b}
\ee
and (\ref{3.23}) yields
\be
s=\pm \int{\frac{\sqrt{1-\frac{1}{N}}d\beta}{\sqrt{N\lambda\beta^{2\left(1-\frac{1}{N}\right)}+2c+c'^2}}} \ .
\label{3.35c}
\ee
In the case (\ref{3.25a}) $\lambda=0$ and $\beta\sim s$. From (\ref{3.35a})-(\ref{3.35c}) one obtains vacuum metrics (1) 
with components depending on one spacelike coordinate $s$ (the same is true
if $\beta=const$, see (\ref{3.24b}) and (\ref{3.24c})). They are not particularly interesting from  point of our method  since they can be obtained by a straightforward integration of the Einstein equations. For
$N=2$ they belong to the class of metrics found by Kasner \cite{K}.

In the case (\ref{3.35}) for $N=2$ equations (\ref{3.35a})-(\ref{3.35c}) can be solved analitycally. 
For $\lambda\neq 0$ one is led to the following
4-dimensional vacuum solution of the Einstein equations
\be\nonumber
g=\epsilon'\left(s+s_0\right)^2du\left(\frac{\lambda
dv}{(1+\frac{\lambda}{4} uv)^2}+\ln{\left|\frac{s-s_0}{s+s_0}\right|} h'(u)du\right)
\ee
\be
-\left|\frac{s+s_0}{s-s_0}\right|ds^2-\left|\frac{s-s_0}{s+s_0}\right|d\varphi^2\ ,\label{3.35d}
\ee
where $\epsilon'=\pm 1$ is the sign of $(s^2-s_0^2)$, $s_0=\frac{c'}{\sqrt{2}\lambda}$ is a constant and $h'=h/s_0$ is an
arbitrary function of $u$. The constant $\lambda$ can be replaced by any nonzero value without a loss of generality. The vector field $\partial_v$ generates a null
geodesic shear-free congruence with no twist and expansion. Metric
(\ref{3.35d}) belongs to a class of the Kundt metrics found
by Kramer and Neugebauer \cite{KN}. This class contains also the following metrics obtained from
(\ref{3.35})-(\ref{3.35c}) for $c=0$ and $\lambda=0$
\beq
g&=&\epsilon'du\left(dv+\ln{\left|s\right|}h(u)du\right)-|s|^{-1}ds^2-\left|s\right|d\phi^2\ ,\label{3.35e}
\eeq
where now $\epsilon'=\pm 1$ is the sign of $s$.

If $\epsilon=-1$, $\Lambda=0$ and $n>1$ simple solutions of (\ref{3.14}) and 
(\ref{3.23}) can be obtained under the assumption that $V$ and $\beta$ have the form $as^b$, where $a$ and $b$ are constants. In this case one obtains $c=0$ and
\be
e^{-2\sqrt{\frac{n+N-1}{n\left(N-1\right)}}\phi}=\frac{(N-1)^2}{(n+N-1)^2}s^{-2}\ ,\ \ \beta=\beta_0 s^{\frac{nN}{n+N-1}}\ ,\ \ \lambda=0 \label{3.26b}
\ee
or
\be
e^{-2\sqrt{\frac{n+N-1}{n\left(N-1\right)}}\phi}=\frac{(N-1)^2}{(n-1)(n+N-1)}s^{-2}\ ,\ \ \beta=\beta_0s^N\ ,\ \lambda=-\frac{(N-1)^2}{n+N-1}\beta_0^{2/N}\ ,\label{3.26a}
\ee
where $\beta_0=const$. Let $r$ be a new coordinate given by
\be
r=s^{\frac{N-1}{n+N-1}}.\label{3.27a}
\ee
After  a minor reparametrization of variables $u$, $v$ and $h$, for $N=2$ one obtains from (\ref{3.35}),    (\ref{3.26a}) and (\ref{3.26b}) the following vacuum metrics
\be
g=du\left(dv-r^{1-n}h(u)du\right)-dr^2-r^2s_{AB}dx^Adx^B\ \label{3.27c}
\ee
\be
g=du\left(\frac{4r^2dv}{(n+1)(1-uv)^2}-r^{1-n}h(u)du\right)-dr^2-\frac{n-1}{n+1}r^2s_{AB}dx^Adx^B\ .\label{3.27b}
\ee
They  are examples of $(n+3)$--dimensional  generalized Kundt metrics \cite{C,CMPP}. In both cases the vector $\partial_v$ defines a  shear- and twist-free congruence of null geodesics.
Metric (\ref{3.27c}) has vanishing scalar invariants \cite{CFHP}. It is particular case of generalized pp wave of type N. As such it can be  easily obtained by  standard methods of general relativity \cite{Kr}. It tends to the Minkowski metric when $r\rightarrow \infty$. Thus, it is asymptotically flat on any timelike section given by $u=u(t),\ v=v(t)$. For instance, if $n=2$ the section $u=v=t$ is 4-dimensional and the corresponding Newton potential takes the form $h(t)/r$. An interpretation of this metric within the brane-world gravity is unclear since the exterior curvature of the  section does not yield any reasonable energy-momentum tensor. 

Metric (\ref{3.27b}) is of type II in generalized Petrov classification \cite{C}. Its Kretschmann scalar $R_{\alpha\beta\gamma\delta}R^{\alpha\beta\gamma\delta}$ nowhere vanishes and it is proportional to $r^{-4}$. Thus, this metric is an example, perhaps the only known explicitly, of a multidimensional Kundt metric with nonconstant scalar invariants (see \cite{CHP} for a discussion of metrics with constant invariants). This metric is singular at $r=0$ and at $uv=1$. The latter singularity can be moved to infinity by means of the transformation
\be
u'=u^{-1}\ ,\ \ \ v'=-\frac{2ur^2}{(n+1)(1-uv)}\label{3.36}
\ee
which puts the metric into the following form
\be
g=du'\left(2dv'-\frac{4v'}{r}dr-\big ((n+1)\frac{v'^2}{r^2}+r^{1-n}h'(u')\big )du'\right)-dr^2-\frac{n-1}{n+1}r^2s_{AB}dx^Adx^B\ .\label{3.37}
\ee

Solution (\ref{3.26b}) can be also merged  with (\ref{3.25a}) for $N>2$ provided ${Tr}{P^2}=0$. In this way one obtains vacuum metrics of the form
\be
g=(\gamma e^{Pr^{1-n}})_{ij}dx^idx^j-dr^2-r^2s_{AB}dx^Adx^B\ .\label{3.38}
\ee
For instance, for N=3, substituting  (\ref{d5}) with $a=\pm\sqrt{3}b$ into (\ref{3.38}) yields the following $(n+4)$-dimensional metric singular at $r=0$
\be
e^{br^{1-n}}\cos{(\sqrt{3}br^{1-n})}(dt^2-dx^2)+2e^{br^{1-n}}\sin{(\sqrt{3}br^{1-n})}dtdx-e^{-2br^{1-n}}dy^2-dr^2-r^2s_{AB}dx^Adx^B.\label{3.39}
\ee

\section{Summary}
We have considered  multidimensional metrics (\ref{1}) invariant under the group $SO(n+1)$ acting on n-dimensional spheres. 
For these metrics, we have reduced vacuum Einstein equations with cosmological constant to lower dimensional Einstein equations with a scalar field. In section 3 we proposed an ansatz which simplifies  these equations for any potential of the scalar field.  Our method is summarized  at the end of section 3. Using this approach in section 4 we were able to rediscover known vacuum solutions of the form (\ref{1}) and to find new ones (see e.g. (\ref{3.27b}) and (\ref{3.38})). Note that equations (\ref{3.14}), (\ref{3.23}) do not depend on details of the matrices $\gamma$ and $P$ except the trace of $P^2$. Thus, it might be possibile to  generalize already known solutions if they satisfy assumptions of our method.

The presented reduction of the Einstein equations is different from that  in brane-world gravity \cite{SMS,G}. In the framework of this theory  our method can be used to find $SO(n+1)$ symmetric bulk metric on one side of a  brane. This metric can be extended to the other side  in such a way that the exterior curvature has a jump corresponding to  matter fields located on the brane (see e.g. examples in \cite{G}). It is highly nontrivial to obtain such a configuration which describes a physically realistic situation (work in progress).

 \null
 
\noindent 
{\bf Acknowledgments}. 
This work was partially supported by the Polish  Committee for Scientific Research (grant 1 PO3B 075 29).

\end{document}